\documentclass
[preprint,letterpaper,noshowpacs,10pt,tightenlines,lengthcheck,onecolumn,prl,balancelastpage]{revtex4}%
\usepackage{amsfonts}
\usepackage{amsmath}
\usepackage{amssymb}
\usepackage{graphicx}%
\begin{document}
\title{Super controlled gates and controlled gates in two-qubit gate simulations}
\author{Ming-Yong Ye}
\email{myye@mail.ustc.edu.cn}
\author{Yong-Sheng Zhang}
\email{yshzhang@ustc.edu.cn}
\author{Guang-Can Guo}
\email{gcguo@ustc.edu.cn}
\affiliation{Laboratory of Quantum Information, Department of Physics, University of
Science and Technology of China (CAS), Hefei 230026, People's Republic of China}

\begin{abstract}
In two-qubit gate simulations an entangling gate is used several times
together with single qubit gates to simulate another two-qubit gate. We show
how a two-qubit gate's simulation power is related to the simulation power of
its mirror gate. And we show that an arbitrary two-qubit gate can be simulated
by three applications of a super controlled gate together with single qubit
gates. We also give the gates set that can be simulated by $n$ applications of
a controlled gate in a constructive way. In addition we give some gates which
can be used four times to simulate an arbitrary two-qubit gate.

PACS number(s): 03.67.Lx, 03.67.Mn

\end{abstract}
\maketitle

We want to make a quantum computer because it can solve some difficult
problems using smaller resource than that needed\ for classical computers. An
\textit{n}-qubit quantum computation can be accomplished by applying an
\textit{n}-qubit gate on a standard initial state followed by a measurement
\cite{nc}. Any \textit{n}-qubit gate can be decomposed as a sequence of the
gates from a universal gate set, which contains all single qubit gates and an
arbitrary two-qubit entangling gate \cite{nc,betal,bddgh,zvsw}. Usually single
qubit gates can be implemented easily and rapidly compared to two-qubit gates,
so it is very important to decompose an \textit{n}-qubit gate by using the
minimum number of a given two-qubit entangling gate in order to exhibit the
power of a quantum computer. Usually to solve this question we need to know
what gates can be simulated by $n\left(  \geq2\right)  $ applications of the
given entangling gate together with single qubit gates. But we do not know
much even for the decomposition of two-qubit gates when the given gate is a
general entangling gate \cite{betal,bddgh,zvsw,vd,control,bgate,vw,sbm}. This
problem is important because our decomposition should rely on the entangling
gate that can be directly generated from the experiment.

In this paper we only investigate the decomposition or simulation of two-qubit
gates. According to the canonical decomposition of two-qubit gate \cite{kc},
we can always write a two-qubit gate in the form: $U_{AB}=\left(  U^{A}\otimes
U^{B}\right)  U_{d}\left(  V^{A}\otimes V^{B}\right)  $, where $U^{A}%
,U^{B},V^{A}$ and $V^{B}$ are single qubit gates and $U_{d}$ has a special
form
\[
U_{d}\left(  \alpha_{1},\alpha_{2},\alpha_{3}\right)  =\exp\left(  i\sum
_{j=1}^{3}\alpha_{j}\sigma_{j}^{A}\otimes\sigma_{j}^{B}\right)  ,
\]
where $\sigma_{1,2,3}$ are Pauli matrix. We can let the parameters satisfy
$\pi/4\geqslant\alpha_{1}\geqslant\alpha_{2}\geqslant\left\vert \alpha
_{3}\right\vert \geqslant0$. The special form $U_{d}$, which is locally
equivalent to $U_{AB}$, is called the canonical form of the two-qubit
gate\ $U_{AB}$.\ The canonical form of the CNOT, DCNOT and SWAP gates are
$U_{d}\left(  \pi/4,0,0\right)  $, $U_{d}\left(  \pi/4,\pi/4,0\right)  $, and
$U_{d}\left(  \pi/4,\pi/4,\pi/4\right)  $ respectively \cite{hvc}. A very
important character of $U_{d}$ is that the magic basis states are its
eigenstates \cite{kc,geometric}, \textit{i.e.} $U_{d}\left(  \alpha_{1}%
,\alpha_{2},\alpha_{3}\right)  \left\vert \Phi_{j}\right\rangle =e^{i\lambda
_{j}}\left\vert \Phi_{j}\right\rangle $, where%

\[
\left\vert \Phi_{1}\right\rangle =1/\sqrt{2}\left(  \left\vert 00\right\rangle
+\left\vert 11\right\rangle \right)  ,\left\vert \Phi_{2}\right\rangle
=i/\sqrt{2}\left(  \left\vert 01\right\rangle +\left\vert 10\right\rangle
\right)  ,
\]%
\[
\left\vert \Phi_{3}\right\rangle =1/\sqrt{2}\left(  \left\vert 01\right\rangle
-\left\vert 10\right\rangle \right)  ,\left\vert \Phi_{4}\right\rangle
=i/\sqrt{2}\left(  \left\vert 00\right\rangle -\left\vert 11\right\rangle
\right)  ,
\]%
\[
\lambda_{1}=+\alpha_{1}-\alpha_{2}+\alpha_{3},\lambda_{2}=+\alpha_{1}%
+\alpha_{2}-\alpha_{3},
\]%
\[
\lambda_{3}=-\alpha_{1}-\alpha_{2}-\alpha_{3},\lambda_{4}=-\alpha_{1}%
+\alpha_{2}+\alpha_{3}.
\]
The controlled gates and super controlled gates in this paper denote the gates
that have canonical forms $U_{d}\left(  \alpha_{1},0,0\right)  $ and
$U_{d}\left(  \pi/4,\alpha_{2},0\right)  $ respectively. In the geometric
representation of two-qubit gates \cite{geometric} controlled gates are
represented by the line between the identity gate and the CNOT gate, and super
controlled gates are represented by the line between the CNOT gate and the
DCNOT gate. Assume that the two-qubit gates $U_{1}$ and $U_{2}$ have canonical
forms $U_{d}\left(  \alpha_{1},\alpha_{2},\alpha_{3}\right)  $ and
$U_{d}\left(  \beta_{1},\beta_{2},\beta_{3}\right)  $ respectively. We say
that the two-qubit gate $U_{2}$ is a mirror gate of $U_{1}$, if $U_{d}\left(
\alpha_{1}+\pi/4,\alpha_{2}+\pi/4,\alpha_{3}+\pi/4\right)  $ is locally
equivalent to $U_{d}\left(  \beta_{1},\beta_{2},\beta_{3}\right)  $. It is not
hard to see that the DCNOT gate is a mirror gate of the CNOT gate.

Now we present a general result about two-qubit simulation.

\textbf{Theorem 1 }(mirror gate theorem) Assume that the gate $U_{d}\left(
\gamma_{1},\gamma_{2},\gamma_{3}\right)  $ is used in a quantum simulation
circuit and the circuit simulates the two-qubit gate $U_{d}\left(  \beta
_{1},\beta_{2},\beta_{3}\right)  $. Then we can replace the gate $U_{d}\left(
\gamma_{1},\gamma_{2},\gamma_{3}\right)  $ in the circuit by one of its mirror
gate to simulate a mirror gate of $U_{d}\left(  \beta_{1},\beta_{2},\beta
_{3}\right)  $.

\textbf{Proof} From the fact that the SWAP gate is locally equivalent to
$U_{d}\left(  \pi/4,\pi/4,\pi/4\right)  $, we can write $U_{SWAP}=s_{1}%
^{A}s_{2}^{B}U_{d}\left(  \pi/4,\pi/4,\pi/4\right)  s_{3}^{A}s_{4}^{B}$ using
some single qubit gates $s_{1}$, $s_{2}$, $s_{3}$, and $s_{4}$. Without loss
of generality, we denote all the gates applied after and before $U_{d}\left(
\gamma_{1},\gamma_{2},\gamma_{3}\right)  $ in the circuit by $k_{1}^{A}%
k_{2}^{B}U_{d}\left(  a_{1},a_{2},a_{3}\right)  k_{3}^{A}k_{4}^{B}$ and
$U_{before}$ respectively, where $\left\{  k_{i}\right\}  _{i=1}^{4}$ is
single qubit gate. Then we have%
\begin{align}
&  k_{1}^{A}k_{2}^{B}U_{d}\left(  a_{1},a_{2},a_{3}\right)  k_{3}^{A}k_{4}%
^{B}U_{d}\left(  \gamma_{1},\gamma_{2},\gamma_{3}\right)  U_{before}%
\nonumber\\
&  =U_{d}\left(  \beta_{1},\beta_{2},\beta_{3}\right)  .
\end{align}
From the Equation (1) we can get%
\begin{align}
&  Sk_{1}^{A}k_{2}^{B}U_{d}\left(  a_{1},a_{2},a_{3}\right)  k_{3}^{A}%
k_{4}^{B}S^{-1}SU_{d}\left(  \gamma_{1},\gamma_{2},\gamma_{3}\right)
U_{before}\nonumber\\
&  =SU_{d}\left(  \beta_{1},\beta_{2},\beta_{3}\right)  ,
\end{align}
where $S=U_{SWAP}\left(  s_{3}^{A}s_{4}^{B}\right)  ^{-1}$. We have two facts
about the two-qubit gate $S$:%
\begin{align}
&  Sk_{1}^{A}k_{2}^{B}U_{d}\left(  a_{1},a_{2},a_{3}\right)  k_{3}^{A}%
k_{4}^{B}S^{-1}\nonumber\\
&  =\left(  s_{4}^{A}s_{3}^{B}\right)  ^{-1}k_{2}^{A}k_{1}^{B}U_{d}\left(
a_{1},a_{2},a_{3}\right)  k_{4}^{A}k_{3}^{B}\left(  s_{4}^{A}s_{3}^{B}\right)
,
\end{align}
and%
\begin{align}
&  SU_{d}\left(  \gamma_{1},\gamma_{2},\gamma_{3}\right) \nonumber\\
&  =s_{1}^{A}s_{2}^{B}U_{d}\left(  \pi/4+\gamma_{1},\pi/4+\gamma_{2}%
,\pi/4+\gamma_{3}\right)  .
\end{align}
Now we rewrite the Equation (2) by using the two facts about $S$ and we have%
\begin{equation}
k_{1}^{A}k_{2}^{B}U_{d}\left(  a_{1},a_{2},a_{3}\right)  k_{3}^{A}k_{4}%
^{B}MU_{before}=N, \label{5}%
\end{equation}
where
\begin{align*}
M  &  =\left(  k_{3}^{A}k_{4}^{B}\right)  ^{-1}k_{4}^{A}k_{3}^{B}\left(
s_{4}^{A}s_{3}^{B}\right)  s_{1}^{A}s_{2}^{B}\\
&  U_{d}\left(  \pi/4+\gamma_{1},\pi/4+\gamma_{2},\pi/4+\gamma_{3}\right)  ,
\end{align*}
and
\begin{align*}
N  &  =k_{1}^{A}k_{2}^{B}\left(  k_{2}^{A}k_{1}^{B}\right)  ^{-1}s_{4}%
^{A}s_{3}^{B}s_{1}^{A}s_{2}^{B}\\
&  U_{d}\left(  \pi/4+\beta_{1},\pi/4+\beta_{2},\pi/4+\beta_{3}\right)  .
\end{align*}
From the Equation (5) we can see that if $U_{d}\left(  \gamma_{1},\gamma
_{2},\gamma_{3}\right)  $ is replaced by its mirror gate $M$ in the circuit,
the simulated gate $U_{d}\left(  \beta_{1},\beta_{2},\beta_{3}\right)  $ is
replaced by its mirror gate $N$.

Some important results can be derived from the mirror gate theorem.

\textbf{Corollary 1} If a two-qubit gate can be simulated by two applications
of $U_{AB}$ together with single qubit gates, then the two qubit gate can also
be simulated by two applications of the mirror gate of $U_{AB}$.

This result is from the fact that the mirror gate's mirror gate is locally
equivalent to the original gate. Since the DCNOT gate is a mirror gate of the
CNOT gate, the gates set that can be simulated by two applications of the
DCNOT gate is the same as that of the two applications of the CNOT gate. This
gates set has been pointed out by Vidal and Dawson \cite{vd}.

\textbf{Corollary 2} If an arbitrary two-qubit gate can be simulated by $n$
applications of $U_{AB}$ together with single qubit gates, then an arbitrary
two-qubit gate can also be simulated by $n$ applications of the mirror gate
of\ $U_{AB}$ \cite{yshzhang}.

It has been showed that an arbitrary two-qubit gate can be simulated by three
applications of the CNOT gate \cite{vd,control,vw,sbm}, so immediately we can
conclude that an arbitrary two-qubit gate can also be simulated by three
applications of the DCNOT gate. This result has been pointed out in
\cite{control}\textit{.}

The B gate can be used two times together with single qubit gates to simulate
an arbitrary two-qubit gate \cite{bgate}. Then it comes to the question: what
gate can be used three times to simulate an arbitrary two-qubit gate?

\textbf{Theorem 2} Three applications of\ the super controlled gate
$U_{d}\left(  \pi/4,\alpha_{2},0\right)  $ together with single qubit gates
can simulate any two-qubit gate $U_{d}\left(  h_{1},h_{2},h_{3}\right)  $. Two
applications of\ the super controlled gate $U_{d}\left(  \pi/4,\alpha
_{2},0\right)  $ together with single qubit gates can simulate any two-qubit
gate $U_{d}\left(  h_{1},h_{2},0\right)  $.

\textbf{Proof} We first define three two-qubit gates $U_{A}$, $U_{B}$, and
$U_{C}$ by their function: $U_{A}\left\vert \Phi_{1}\right\rangle =\left\vert
00\right\rangle ,$ $U_{A}\left\vert \Phi_{2}\right\rangle =\left\vert
01\right\rangle ,$ $U_{A}\left\vert \Phi_{3}\right\rangle =\left\vert
10\right\rangle ,$ $U_{A}\left\vert \Phi_{4}\right\rangle =e^{4i\alpha_{2}%
}\left\vert 11\right\rangle $, $U_{B}\left\vert 00\right\rangle =\left(
\cos2\alpha_{2}\left\vert 0\right\rangle +\sin2\alpha_{2}\left\vert
1\right\rangle \right)  \left\vert 0\right\rangle $, $U_{B}\left\vert
01\right\rangle =\left\vert 01\right\rangle $, $U_{B}\left\vert
10\right\rangle =e^{8i\alpha_{2}}\left\vert 11\right\rangle $, $U_{B}%
\left\vert 11\right\rangle =\left(  \sin2\alpha_{2}\left\vert 0\right\rangle
-\cos2\alpha_{2}\left\vert 1\right\rangle \right)  \left\vert 0\right\rangle
$, $U_{C}\left(  \cos2\alpha_{2}\left\vert 0\right\rangle +\sin2\alpha
_{2}\left\vert 1\right\rangle \right)  \left\vert 0\right\rangle =\left\vert
\Phi_{1}\right\rangle ,$ $U_{C}\left\vert 01\right\rangle =\left\vert \Phi
_{2}\right\rangle ,$ $U_{C}\left\vert 11\right\rangle =e^{-4i\alpha_{2}%
}\left\vert \Phi_{3}\right\rangle ,$ and $U_{C}\left(  \sin2\alpha
_{2}\left\vert 0\right\rangle -\cos2\alpha_{2}\left\vert 1\right\rangle
\right)  \left\vert 0\right\rangle =\left\vert \Phi_{4}\right\rangle $. We can
verified that the gate $e^{-2ic_{2}}U_{C}e^{ih_{3}\sigma_{3}^{B}}%
U_{B}e^{i\left(  h_{1}+2c_{2}\right)  \sigma_{3}^{A}}e^{-ih_{2}\sigma_{3}^{B}%
}U_{A}$ is just the same gate as $U_{d}\left(  h_{1},h_{2},h_{3}\right)  $ by
applying them on the magic basis states. It can also be verified that
$U_{A}^{-1}e^{ih_{1}\sigma_{3}^{A}}e^{-ih_{2}\sigma_{3}^{B}}U_{A}$ is the same
gate as $U_{d}\left(  h_{1},h_{2},0\right)  $. Using the methods given in
\cite{kc,geometric}, we can find that $U_{A}$, $U_{B}$, $U_{C}$, and
$U_{A}^{-1}$ are locally equivalent to the super controlled gate $U_{d}\left(
\pi/4,\alpha_{2},0\right)  $. So we can end our proof.

Now we turn to investigate the gates set that can be simulated by $n$
applications of a controlled gate.

\textbf{Theorem 3 }Given two controlled gates $e^{\frac{i}{2}\gamma_{1}%
\sigma_{3}^{A}\otimes\sigma_{3}^{B}}$ and $e^{\frac{i}{2}\gamma_{2}\sigma
_{3}^{A}\otimes\sigma_{3}^{B}}$ with $0<\gamma_{2}\leq\gamma_{1}\leq\pi/2$. If
a two-qubit gate that can be simulated by these two controlled gates together
with single qubit gates, it must be locally equivalent to a gate $U_{d}\left(
h_{1}/2,h_{2}/2,0\right)  $ with $0\leq h_{2}\leq h_{1}\leq\pi/2$, $\gamma
_{1}-\gamma_{2}\leq h_{1}-h_{2}$, and $h_{1}+h_{2}\leq\gamma_{1}+\gamma_{2}$.

This problem has been investigated by \textit{Zhang et al}. \cite{control},
but their result is incomplete. We will give the proof later. From
\textbf{theorem 3} we can derive the following result.

\textbf{Corollary 3} The two-qubit gate $U_{d}\left(  h_{1}/2,h_{2}%
/2,0\right)  $ with $\left\vert h_{1}\right\vert +\left\vert h_{2}\right\vert
\leq n\gamma$, can be simulated by $n\left(  \geq2\right)  $ applications of
the controlled gate $e^{\frac{i}{2}\gamma\sigma_{3}^{A}\otimes\sigma_{3}^{B}}$
together with single qubit gates, where $0<\gamma\leq\pi/2$.

\textbf{Proof} We only need to prove the result when $0\leq h_{2}\leq
h_{1}\leq\pi/2$, and $h_{1}+h_{2}\leq n\gamma$, because $U_{d}\left(
h_{1}/2,h_{2}/2,0\right)  $ is locally equivalent to a gate $U_{d}\left(
h_{1}^{\prime}/2,h_{2}^{\prime}/2,0\right)  $ with $0\leq h_{2}^{\prime}\leq
h_{1}^{\prime}\leq\pi/2$, and $h_{1}^{\prime}+h_{2}^{\prime}\leq n\gamma$.
When $n=2$ the corollary is obvious from \textbf{theorem 3}. Assume that the
corollary is true for $n=m\geq2$, we only need to prove that the corollary is
also true for $n=m+1$. Because two applications of the controlled gate
$e^{\frac{i}{2}\gamma\sigma_{3}^{A}\otimes\sigma_{3}^{B}}$ can simulate
itself, the gate which can be simulated by $n$ applications of the controlled
gate can also be simulated by $n+1$ applications of the controlled gate. So we
only need to prove that the gate $U_{d}\left(  h_{1}/2,h_{2}/2,0\right)  $
with $0\leq h_{2}\leq h_{1}\leq\pi/2$ and $m\gamma\leq h_{1}+h_{2}\leq\left(
m+1\right)  \gamma$, can be simulated by $m+1$ applications of the controlled
gate $e^{\frac{i}{2}\gamma\sigma_{3}^{A}\otimes\sigma_{3}^{B}}$ together with
single qubit gates. It is not hard to find that the gate $U_{d}\left(
h_{1}/2,h_{2}/2,0\right)  $ can be simulated by $U_{d}\left(  \left(
h_{1}-\gamma\right)  /2,h_{2}/2,0\right)  $ and $e^{\frac{i}{2}\gamma
\sigma_{3}^{A}\otimes\sigma_{3}^{B}}$ together with single qubit gates. So we
only need to prove that the gate $U_{d}\left(  \left(  h_{1}-\gamma\right)
/2,h_{2}/2,0\right)  $ can be simulated by $m$ applications of the controlled
gate $e^{\frac{i}{2}\gamma\sigma_{3}^{A}\otimes\sigma_{3}^{B}}$. This is true
since both $h_{1}-\gamma$ and $h_{2}$ are in the interval $\left[
0,\pi/2\right]  $ and $h_{1}-\gamma+h_{2}\leq m\gamma$. So we can end our proof.

Zhang \textit{et. al. }\cite{control}\textit{ }have given the gates set that
can be simulated by $n$ $\left(  n\geqslant3\right)  $ applications of a
controlled-unitary gate, but they have not given the explicit simulation
method. Here we give an explicit simulation method for $n\geqslant4$.

\textbf{Theorem 4} The gate $U_{d}\left(  h_{1}/2,h_{2}/2,h_{3}/2\right)  $
with $0\leq\left\vert h_{3}\right\vert \leq h_{2}\leq h_{1}\leq\pi/2$, can be
simulated by $n$ $\left(  n\geq3\right)  $ applications of the gate
$e^{\frac{i}{2}\gamma\sigma_{3}^{A}\otimes\sigma_{3}^{B}}$ together with
single qubit gates if $h_{1}+h_{2}+\left\vert h_{3}\right\vert \leq n\gamma$,
where $0<\gamma\leq\pi/2$.

\textbf{Proof} This result is proved in \cite{control}, but their proof is not
constructive. Here we give a constructive procedure to simulate the gates by
$n\left(  n\geq4\right)  $ applications of the controlled gate $e^{\frac{i}%
{2}\gamma\sigma_{3}^{A}\otimes\sigma_{3}^{B}}$. First we have $\left\vert
h_{3}\right\vert \leq n\gamma/3$. We denote $m=\left\lceil n/3\right\rceil $,
the function $\left\lceil x\right\rceil $ is defined as the smallest integer
which is not smaller that $x$. When $n\geq4$, we have $n-m\geq m\geq2$. Based
on \textbf{corollary 3 }we can find that the gate $e^{\frac{i}{2}\left(
\left(  m\gamma-\left\vert h_{3}\right\vert \right)  \sigma_{1}^{A}%
\otimes\sigma_{1}^{B}+h_{3}\sigma_{3}^{A}\otimes\sigma_{3}^{B}\right)  }$ can
be simulated by $m$ applications of the controlled gate $e^{\frac{i}{2}%
\gamma\sigma_{3}^{A}\otimes\sigma_{3}^{B}}$, because $\left\vert
m\gamma-\left\vert h_{3}\right\vert \right\vert +\left\vert h_{3}\right\vert
=m\gamma$. Similarly we can find that the gate $e^{\frac{i}{2}\left(  \left(
h1-m\gamma+\left\vert h_{3}\right\vert \right)  \sigma_{1}^{A}\otimes
\sigma_{1}^{B}+h_{2}\sigma_{2}^{A}\otimes\sigma_{2}^{B}\right)  }$ can be
simulated by $\left(  n-m\right)  $ applications of the controlled gate
$e^{\frac{i}{2}\gamma\sigma_{3}^{A}\otimes\sigma_{3}^{B}}$, because
$\left\vert h_{1}-m\gamma+\left\vert h_{3}\right\vert \right\vert +h_{2}%
\leq\left(  n-m\right)  \gamma$. Notice that the gate $\exp\left(  \frac{i}%
{2}\sum_{j=1}^{2}h_{j}\sigma_{j}^{A}\otimes\sigma_{j}^{B}\right)  $ is just
the product of $e^{\frac{i}{2}\left(  \left(  m\gamma-\left\vert
h_{3}\right\vert \right)  \sigma_{1}^{A}\otimes\sigma_{1}^{B}+h_{3}\sigma
_{3}^{A}\otimes\sigma_{3}^{B}\right)  }$ and $e^{\frac{i}{2}\left(  \left(
h1-m\gamma+\left\vert h_{3}\right\vert \right)  \sigma_{1}^{A}\otimes
\sigma_{1}^{B}+h_{2}\sigma_{2}^{A}\otimes\sigma_{2}^{B}\right)  }$. So we can
end our proof.

The condition $h_{1}+h_{2}+\left\vert h_{3}\right\vert \leq n\gamma$ is also a
necessary condition for the gate that can be simulated by $n$ $\left(
n\geq3\right)  $ applications of the gate $e^{\frac{i}{2}\gamma\sigma_{3}%
^{A}\otimes\sigma_{3}^{B}}$ together with single qubit gates
\cite{control,yshzhang}. \textbf{Theorem 3} and \textbf{theorem 4} tell us the
gates set that can be simulated by $n$ $\left(  \geq2\right)  $ applications
of the controlled gate. It is not hard to find out that $\left\lceil
\frac{3\pi}{2\gamma}\right\rceil $ applications of the controlled gate
$e^{\frac{i}{2}\gamma\sigma_{3}^{A}\otimes\sigma_{3}^{B}}$ can simulate an
arbitrary two-qubit gate. According to the mirror gate theorem, we can easily
find out the gates set that can be simulated by $n$ $\left(  \geq2\right)  $
applications of the gate $U_{d}\left(  \pi/4,\pi/4,\pi/4+\gamma/2\right)  $.
And we can conclude that $\left\lceil \frac{3\pi}{2\gamma}\right\rceil $
applications of the gate $U_{d}\left(  \pi/4,\pi/4,\pi/4+\gamma/2\right)  $
together with single qubit gates can simulate an arbitrary two-qubit gate.

We can only give part results for $n=3$ in \textbf{theorem 4}. We first prove
the following result.

\textbf{Theorem 5} The two-qubit gate
\begin{align*}
U_{AB}  &  =U_{d}\left(  a_{1}/2,a_{2}/2,a_{3}/2\right) \\
&  \left(  e^{\frac{^{i}}{2}s_{1}\sigma_{y}^{A}}\otimes e^{\frac{^{i}}{2}%
s_{2}\sigma_{y}^{B}}\right)  U_{d}\left(  b_{1}/2,b_{2}/2,b_{3}/2\right)
\end{align*}
is locally equivalent to the gate $U_{d}\left(  x/2,y/2,\left(  a_{2}%
+b_{2}\right)  /2\right)  $, where
\begin{align*}
\cos\left(  x+y\right)   &  =\cos\left(  a_{1}+a_{3}\right)  \cos\left(
b_{1}+b_{3}\right) \\
&  -\cos\left(  s_{1}-s_{2}\right)  \sin\left(  a_{1}+a_{3}\right)
\sin\left(  b_{1}+b_{3}\right)  ,
\end{align*}
and
\begin{align*}
\cos\left(  x-y\right)   &  =\cos\left(  a_{1}-a_{3}\right)  \cos\left(
b_{1}-b_{3}\right) \\
&  -\cos\left(  s_{1}+s_{2}\right)  \sin\left(  a_{1}-a_{3}\right)
\sin\left(  b_{1}-b_{3}\right)  .
\end{align*}

\textbf{Proof }Following the procedure in \cite{kc,geometric}, we first write
the gate $U_{AB}$ in the magic basis. The gate $U_{AB}^{T}$ represents the
transpose of $U_{AB}$ in the magic basis. Notice that $\left(  e^{\frac{^{i}%
}{2}s_{1}\sigma_{y}^{A}}\otimes e^{\frac{^{i}}{2}s_{2}\sigma_{y}^{B}}\right)
$ can be regarded as a block diagnosed matrix in the magic basis, It is not
hard to find that both the gate $U_{AB}$ and $M=U_{AB}^{T}U_{AB}$ can be
regarded as block diagnosed matrix in the magic basis. As usual, we just
compute the eigenvalues of the matrix $M$ to find the canonical form of the
gate $U_{AB}$. Because the gate $M$ is represented by a block diagnosed matrix
in the magic basis, we can easily find that the four eigenvalues of $M$ have
the following relations:%
\begin{align*}
x_{1}x_{2}  &  =e^{-2i\left(  a_{2}+b_{2}\right)  },x_{3}x_{4}=e^{2i\left(
a_{2}+b_{2}\right)  },\\
x_{1}+x_{2}  &  =2e^{-i\left(  a_{2}+b_{2}\right)  }\cos\left(  x+y\right)
,\\
x_{3}+x_{4}  &  =2e^{i\left(  a_{2}+b_{2}\right)  }\cos\left(  x-y\right)  ,
\end{align*}
where $\cos\left(  x+y\right)  $ and $\cos\left(  x-y\right)  $ are given in
the theorem. So we can write $x_{1}=e^{-i\left(  a_{2}+b_{2}\right)
}e^{i\left(  x+y\right)  }$, $x_{2}=e^{-i\left(  a_{2}+b_{2}\right)
}e^{-i\left(  x+y\right)  }$, $x_{3}=e^{i\left(  a_{2}+b_{2}\right)
}e^{i\left(  x-y\right)  }$, $x_{4}=e^{i\left(  a_{2}+b_{2}\right)
}e^{-i\left(  x-y\right)  }$. Compare with the eigenvalues of $M$ computed
from a two-qubit gate in the canonical form, without loss of generality we
think $U_{AB}$ is locally equivalent to the gate $U_{d}\left(  x/2,y/2,\left(
a_{2}+b_{2}\right)  /2\right)  $.

\textbf{Theorem 6} The gate $U_{d}\left(  h_{1}/2,h_{2}/2,h_{3}/2\right)  $
can be simulated by $3$ applications of the gate $e^{\frac{i}{2}\gamma
\sigma_{3}^{A}\otimes\sigma_{3}^{B}}$ together with single qubit gates if
$0\leq h_{1}-h_{2}\leq\min\left(  3\gamma-\left\vert h_{3}\right\vert
,\pi\right)  $, $0\leq h_{1}+h_{2}\leq\min\left(  3\gamma-\left\vert
h_{3}\right\vert ,\pi\right)  $, and $\left\vert h_{2}\right\vert \leq\gamma$,
where $0<\gamma\leq\pi/2$.

\textbf{Proof} We first simulate the gate $U_{d}\left(  c_{1}/2,h_{3}%
/2,0\right)  $ with $\gamma\leq c_{1}\leq\min\left(  2\gamma-\left\vert
h_{3}\right\vert ,\pi-\gamma\right)  $ by two applications of the controlled
gate $e^{\frac{i}{2}\gamma\sigma_{3}^{A}\otimes\sigma_{3}^{B}\text{ }}$ based
on \textbf{corollary 3}. Notice that $U_{d}\left(  c_{1}/2,h_{3}/2,0\right)
\left(  e^{\frac{^{i}}{2}s_{1}\sigma_{y}^{A}}\otimes e^{\frac{^{i}}{2}%
s_{2}\sigma_{y}^{B}}\right)  U_{d}\left(  0,0,\gamma/2\right)  $ is locally
equivalent to the gate $U_{d}\left(  h_{1}/2,h_{2}/2,h_{3}/2\right)  $ with
\begin{align*}
\cos\left(  h_{1}+h_{2}\right)   &  =\cos\left(  c_{1}\right)  \cos\left(
\gamma\right) \\
&  -\cos\left(  s_{1}-s_{2}\right)  \sin\left(  c_{1}\right)  \sin\left(
\gamma\right)  ,
\end{align*}
and
\begin{align*}
\cos\left(  h_{1}-h_{2}\right)   &  =\cos\left(  c_{1}\right)  \cos\left(
\gamma\right) \\
&  +\cos\left(  s_{1}+s_{2}\right)  \sin\left(  c_{1}\right)  \sin\left(
\gamma\right)  .
\end{align*}
From the above two equations, we can find that both $\cos\left(  h_{1}%
+h_{2}\right)  $ and $\cos\left(  h_{1}-h_{2}\right)  $ can be any value in
the interval $\left[  \cos\left(  c_{1}+\gamma\right)  ,\cos\left(
c_{1}-\gamma\right)  \right]  $. So we have $c_{1}-\gamma\leq h_{1}-h_{2}\leq
c_{1}+\gamma$ and $c_{1}-\gamma\leq h_{1}+h_{2}\leq c_{1}+\gamma$. When we
vary $c_{1}$ from $\gamma$ to $\min\left(  2\gamma-\left\vert h_{3}\right\vert
,\pi-\gamma\right)  $, we can find that $h_{1}$ and $h_{2}$ can be any value
satisfying conditions: $0\leq h_{1}-h_{2}\leq\min\left(  3\gamma-\left\vert
h_{3}\right\vert ,\pi\right)  $, $0\leq h_{1}+h_{2}\leq\min\left(
3\gamma-\left\vert h_{3}\right\vert ,\pi\right)  $, and $\left\vert
h_{2}\right\vert \leq\gamma$.

Based on \textbf{theorem 5} we can give a simple proof for \textbf{theorem 3}.
Assume $a_{1}=a_{2}=b_{1}=b_{2}=0$, and $0<b_{3}\leq a_{3}\leq\pi/2$ in
\textbf{theorem 5}. Then the gate $e^{\frac{i}{2}a_{3}\sigma_{3}^{A}%
\otimes\sigma_{3}^{B}}\left(  e^{\frac{^{i}}{2}s_{1}\sigma_{y}^{A}}\otimes
e^{\frac{^{i}}{2}s_{2}\sigma_{y}^{B}}\right)  e^{\frac{i}{2}b_{3}\sigma
_{3}^{A}\otimes\sigma_{3}^{B}}$ is locally equivalent to the gate
$U_{d}\left(  x/2,y/2,0\right)  $ with $0\leq y\leq x\leq\pi/2$, where
\begin{align*}
\cos\left(  x+y\right)   &  =\cos\left(  a_{3}\right)  \cos\left(
b_{3}\right) \\
&  -\cos\left(  s_{1}-s_{2}\right)  \sin\left(  a_{3}\right)  \sin\left(
b_{3}\right)  ,
\end{align*}
and
\begin{align*}
\cos\left(  x-y\right)   &  =\cos\left(  a_{3}\right)  \cos\left(
b_{3}\right) \\
&  -\cos\left(  s_{1}+s_{2}\right)  \sin\left(  a_{3}\right)  \sin\left(
b_{3}\right)  .
\end{align*}
From the above two equations, we have $a_{3}-b_{3}\leq x-y$ and $x+y\leq
a_{3}+b_{3}$. According to ZYZ decomposition of single qubit gate \cite{nc},
$e^{\frac{i}{2}a_{3}\sigma_{3}^{A}\otimes\sigma_{3}^{B}}\left(  e^{\frac{^{i}%
}{2}s_{1}\sigma_{y}^{A}}\otimes e^{\frac{^{i}}{2}s_{2}\sigma_{y}^{B}}\right)
e^{\frac{i}{2}b_{3}\sigma_{3}^{A}\otimes\sigma_{3}^{B}}$ can be regarded as a
representative of the gates that can be simulated by $e^{\frac{i}{2}%
a_{3}\sigma_{3}^{A}\otimes\sigma_{3}^{B}}$ and $e^{\frac{i}{2}b_{3}\sigma
_{3}^{A}\otimes\sigma_{3}^{B}}$. So we can end the proof of \textbf{theorem 3}.

In theorem 2 we have shown that three applications of a super controlled gate
can simulate an arbitrary two-qubit gate, but it is still an open question to
find out all the two-qubit gates that have the same simulation power as super
controlled gates. Now we go on to find out some gates, which can be used four
times to simulate any two-qubit gates. Since we know that two applications the
gate $B=U_{d}\left(  \pi/4,\pi/8,0\right)  $ can simulate any two-qubit gate
\cite{bgate}, \ then four applications of the gate $U_{AB}$ can also simulate
any two-qubit gate if two applications of $U_{AB}$ can simulate the gate $B$.

\textbf{Theorem 7} The gate $U_{1}=U_{d}\left(  a_{1}/2,0,a_{3}/2\right)  $
can be used four times to simulate any two-qubit gate if $\cos\left(
2a_{1}+2a_{3}\right)  \leq-1/\sqrt{2}$ and $\cos\left(  2a_{1}-2a_{3}\right)
\leq1/\sqrt{2}$.

\textbf{Proof} We only prove two applications of the gate $U_{1}$ can simulate
the gate B. $U_{2}=U_{d}\left(  a_{3}/2,0,a_{1}/2\right)  $ is locally
equivalent to $U_{1}$. We assume that $U_{1}\left(  e^{\frac{^{i}}{2}%
s_{1}\sigma_{y}^{A}}\otimes e^{\frac{^{i}}{2}s_{2}\sigma_{y}^{B}}\right)
U_{2}$ is locally equivalent to the gate $B$. Then according to
\textbf{theorem 5} we have
\begin{align*}
\cos\left(  \pi/2+\pi/4\right)   &  =\cos^{2}\left(  a_{1}+a_{3}\right)  \\
&  -\cos\left(  s_{1}-s_{2}\right)  \sin^{2}\left(  a_{1}+a_{3}\right)  ,
\end{align*}
and
\begin{align*}
\cos\left(  \pi/2-\pi/4\right)   &  =\cos^{2}\left(  a_{1}-a_{3}\right)  \\
&  +\cos\left(  s_{1}+s_{2}\right)  \sin^{2}\left(  a_{1}-a_{3}\right)  .
\end{align*}
To ensure we can find suitable parameters $s_{1}$ and $s_{2}$ satisfying the
above two equations, we only need%
\begin{align*}
\cos\left(  2a_{1}+2a_{3}\right)   &  \leq\cos\left(  \pi/2+\pi/4\right)
=-1/\sqrt{2},\\
\cos\left(  2a_{1}-2a_{3}\right)   &  \leq\cos\left(  \pi/2-\pi/4\right)
=1/\sqrt{2}.
\end{align*}
So every gate, which is locally equivalent to $U_{d}\left(  a_{1}%
/2,0,a_{3}/2\right)  $, can be used four times to simulate any two-qubit gate
if $\cos\left(  2a_{1}+2a_{3}\right)  \leq-1/\sqrt{2}$ and $\cos\left(
2a_{1}-2a_{3}\right)  \leq1/\sqrt{2}$.

In conclusion, we have given a general result about two-qubit gate simulations
and we have shown that some gates can be used three times or four times to
simulate an arbitrary two-qubit gate. We also give the gates set that can be
simulated by $n$ applications of a controlled gate through a constructive
procedure. These results are important for quantum computer designers to
exhibit the power of quantum computers since their design should base on the
entangling gate that can generate directly from the experiment. The
\textbf{mirror gate theorem} we present gives us another way to find the
simulation power of entangling gates. This result may give some new insight
into gate simulations.

This work was funded by National Fundamental Research Program (Program No.
2001CB309300), National Natural Science Foundation of China (No. 10304017),
and Chinese Innovation Fund (No. Grant 60121503).

\end{document}